\documentclass{article}
\usepackage{amsmath}
\usepackage{amssymb}
\usepackage{geometry}
\usepackage{lineno}
\usepackage{setspace}
\usepackage{color}
\usepackage{graphicx}
\usepackage[hyphens]{url}
\usepackage{natbib}
\usepackage{bm}
\usepackage{xcolor}

\newcommand{\var}{\textrm{var}}
\newcommand{\cov}{\textrm{cov}}
\newcommand{\se}{\textrm{se}}

\begin{document}

\noindent \textbf{Title:}  Comparing Spatial Regression to Random Forests for Large Environmental Data Sets\\

\noindent \textbf{Authors:} Eric W. Fox$^1$, Jay M. Ver Hoef$^2$, and Anthony R. Olsen$^3$\\

\noindent $^1$California State University East Bay, Department of Statistics and Biostatistics, Hayward, CA 94542.\\
\noindent Email: eric.fox@csueastbay.edu\\

\noindent $^2$National Marine Mammal Laboratory, Alaska Fisheries Science Center, National Oceanic and Atmospheric Administration, Seattle, WA 98115.\\

\noindent $^3$National Health and Environmental Effects Research Laboratory, Western Ecology Division, U.S. Environmental Protection Agency, Corvallis, OR 97333.
\clearpage

\begin{abstract}
Environmental data may be ``large" due to number of records, number of covariates, or both.  Random forests has a reputation for good predictive performance when using many covariates with nonlinear relationships, whereas spatial regression, when using reduced rank methods, has a reputation for good predictive performance when using many records that are spatially autocorrelated.  In this study, we compare these two techniques using a data set containing the macroinvertebrate multimetric index (MMI) at 1859 stream sites with over 200 landscape covariates.  A primary application is mapping MMI predictions and prediction errors at 1.1 million perennial stream reaches across the conterminous United States.  For the spatial regression model, we develop a novel transformation procedure that estimates Box-Cox transformations to linearize covariate relationships and handles possibly zero-inflated covariates. We find that the spatial regression model with transformations, and a subsequent selection of significant covariates, has cross-validation performance slightly better than random forests.  We also find that prediction interval coverage is close to nominal for each method, but that spatial regression prediction intervals tend to be narrower and have less variability than quantile regression forest prediction intervals.  A simulation study is used to generalize results and clarify advantages of each modeling approach.\\
\end{abstract}

\clearpage

\section{Introduction}

As we enter the age of ``big data'' \citep{mcafee2012big,lohr2012age}, innovative statistical methods are required for insights from massive data sets \citep{gandomi2015beyond}.  While big data is an abstract concept \citep{chen2014big}, data for a statistical analysis are generally prepared as tables, with records down the rows, and variables across the columns \citep{hand1999statistics}. While we immediately recognize big data problems for large numbers of records, there are also issues when there are large numbers of columns \citep{tukey1991use}.  We are interested in a spatial data set in the United States of national importance based on an aquatic health index, where there are thousands of rows and hundreds of columns.  Two leading candidates for analyzing such data are random forests (\citealp{breiman01rf}), because it handles large numbers of covariates with nonlinear relationships, and spatial regression \citep{verhoef01slm}, because it accounts for possible spatial autocorrelation.  The ultimate goal of the analysis is to predict the aquatic health index at over one million watersheds throughout the country, while also gaining some understanding of which covariates are important.  The goal of this article is to use best practices for analyzing the data with spatial regression and random forests, and then compare the methods.

\subsection{Motivating Data Set} \label{sec:data}
\begin{figure}[htbp]
\centering
\includegraphics[width=0.95\textwidth]{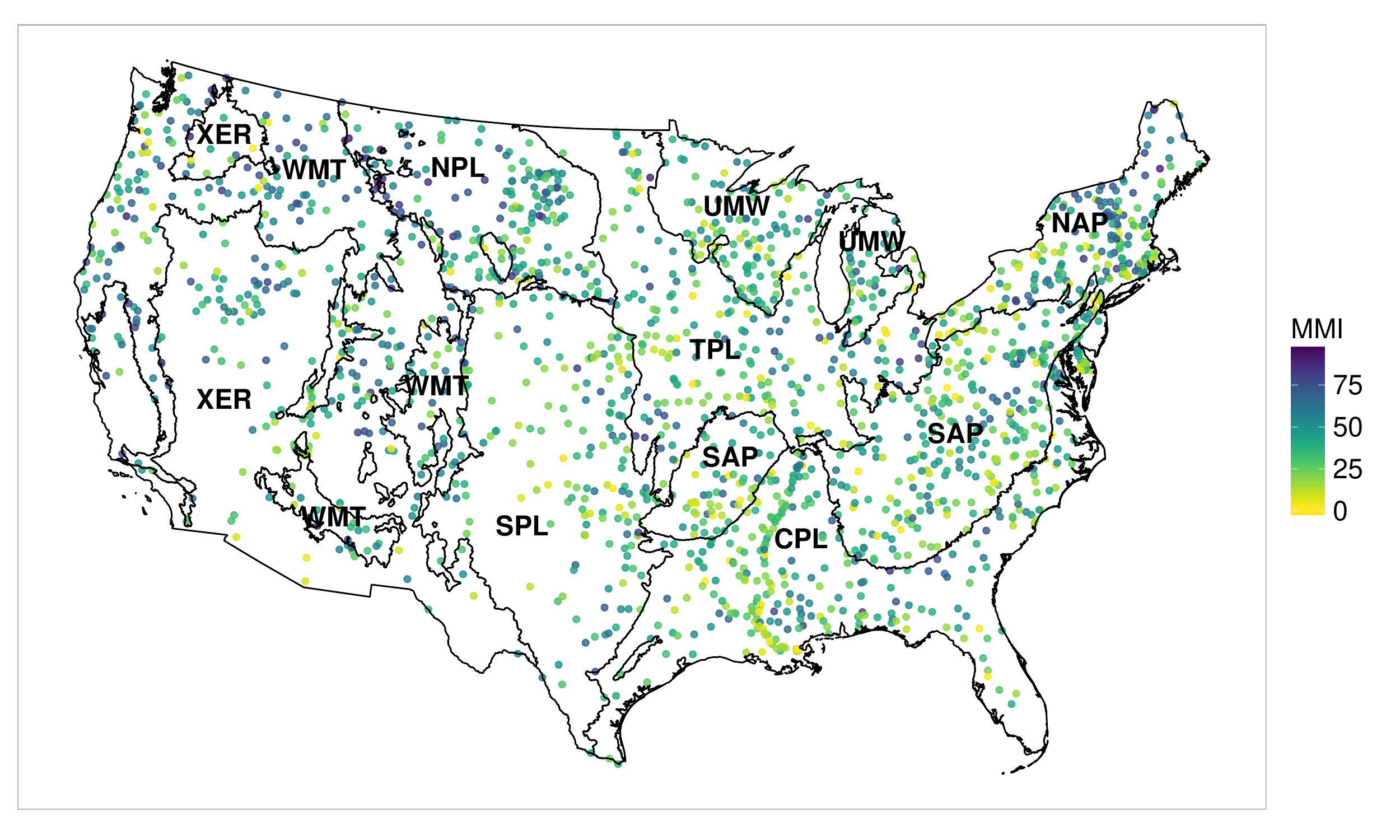}
\caption{Locations of stream sites from the 2008/09 NRSA with point colors corresponding to sampled MMI scores. Ecoregions: Coastal Plains (CPL), Northern Appalachians (NAP), Northern Plains (NPL), Southern Appalachians (SAP), Southern Plains (SPL), Temperate Plains (TPL), Upper Midwest (UMW), Western Mountains (WMT), and Xeric (XER).  \label{fig:nrsa_map}}
\end{figure}

For this study, we use a data set containing the biological condition of $n=1859$ sites from the US Environmental Protection Agency's 2008/09 National Rivers and Streams Assessment (NRSA; \citealp{epa09nrsa}).  The NRSA used a generalized random-tessellation design \citep{olsen2004jasa} to collect a spatially-balanced and representative sample of stream sites across the conterminous US (CONUS).  The target population for the design consisted of all rivers and streams within the CONUS that had flowing water during the study period, which extended between April to September of 2008/09.  Benthic macroinverabrates (e.g., aquatic insects, crustaceans, and worms) were sampled to determine the biological condition of stream sites.  A multimetric index (MMI)  was developed for the NRSA to summarize several measures of the condition of macroinvertebrate assemblages (e.g., taxonomic composition, diversity, tolerance to disturbance, etc.) into a combined index.  The reported MMI values were calculated by summing six individual measures, or `metrics', and then normalized to a 0-100 scale \citep{epa09nrsatech}.  The individual metrics used for the MMI were selected separately for each of the nine ecoregions \citep{omernik87ecoregions} to account for some of the natural variation in climate, geology, hydrology and soils among stream sites.  The ecoregion boundaries and locations of the 2008/09 NRSA stream sites are shown in Figure~\ref{fig:nrsa_map}.  A comprehensive description of the MMI developed for the NRSA was provided by \cite{stoddard08mmi}, \cite{epa09nrsatech}, and \cite{hill17map}.

For modeling the MMI we use a large suite of $p=209$ covariates from the Stream-Catchment (StreamCat) data set (\citealp{hill16streamcat}; publicly available at \url{https://www.epa.gov/national-aquatic-resource-surveys/streamcat}).  StreamCat contains upstream landscape features (e.g., topograpy, precipitation, landscape imperviousness, urban and agricultural land use) for 2.6 million stream reaches across the CONUS and allows for spatially explicit prediction.  The variables in StreamCat can be linked to the National Hydrography Dataset Plus Version 2 (NHDPlusV2; \citealp{mckay2012nhdplus}) and are available at two spatial scales: local catchment and full contributing watershed.  \cite{hill16streamcat} defines a ``catchment" as the local drainage area for an individual NHDPlusV2 stream segement, excluding upstream drainage area; and a ``watershed" as the set of hydrologically connected catchments that contribute flow to a given catchment (i.e., catchment plus upstream catchments).  Note that we only make MMI predictions for the 1.1 million perennial stream reaches (as designated in NHDPlusV2) since the sample frame for the NRSA is limited to these types of streams. Descriptions of StreamCat covariates used in this study are provided in the Supplement.

\subsection{Literature Review} \label{sec:lit}

There have been surprisingly few attempts to compare random forests and spatial regression. Most found random forests superior to various forms of linear regression with autocorrelated errors \citep{li2011application,li2011can,appelhans2015evaluating,hengl2015mapping,fayad2016regional}, although \citet{parmentier2011predicting}, and \citet{temesgen2014evaluation} found spatial regression outperformed random forests.  All comparisons used either root-mean-squared prediction error (RMSPE) or mean absolute prediction error (MAPE) on some form of n-fold cross-validation, and only \citet{temesgen2014evaluation} evaluated the estimated prediction errors to examine whether prediction intervals contained the true values with the correct proportion.  None of the papers simulated spatially autocorrelated data to evaluate and compare the different modeling approaches.

The comparisons in our review used random forests and spatial regression mostly as black box methods.  By black box methods, we mean that data are used without much examination, and methods that rely on default values and little user interaction.  Generally, random forests seems to outperform spatial regression, but only as a black box method.  Can practitioners with more experience make each of these methods work better, and then how will they compare?  It will be (or should be) rare that data, collected at great expense, are subjected to black box methods.  The history of regression has taught statisticians to use best practices for their data analyses, including exploring data, making transformations, checking residuals, using model diagnostics, and then possibly refitting models.  We suggest that the results given in the previously mentioned literature are not necessarily reflective of a considered approach to many data analyses.  In contrast, we will confine ourselves to a single data set, and a single response variable, but we take considerable effort to make each method work as well as possible, and discuss the ramifications after the analyses.  We will also use simulations to investigate properties not seen in the real data.

Our objectives are to compare random forest and spatial regression modeling approaches for predicting and mapping the MMI for all $1.1$ million perennial stream reaches across the CONUS.  In a related study, \cite{hill17map} used random forest modeling to predict the binary ``good" and ``poor" MMI condition classes with StreamCat predictor variables.  The random forest models developed in \cite{hill17map} were used to map the predicted probability of good stream condition for all perennial CONUS stream reaches.  In this article we instead model the MMI scores directly, and include both random forests and spatial regression.  We also evaluate each method's ability to quantify the uncertainty of the MMI predictions.  Previous studies have produced maps of random forest model uncertainties by interpolating the residuals \citep{oliveira2012fire, appelhans2015evaluating}, or taking the standard deviations of the predictions made by each tree in the ensemble \citep{freeman2015tuning}.  In this article we take a different approach, and formally construct random forest prediction intervals using the method of quantile regression forests \citep{meinshausen2006quantile},  which has been studied primarily in the context of non-spatial data.  We also consider a hybrid random forest regression-kriging approach, in which a simple-kriging model is estimated for the random forest residuals, and simple-kriging predictions of residuals are added to random forest predictions.  Although we focus on a particular data set, we generalize the concepts through simulations, and our overall goal is application to other large environmental data sets.

\section{Spatial Regression Model} \label{sec:SLMreview}
Here, we introduce the spatial regression model, likelihood-based estimation methods, and kriging prediction and variance equations that we apply to the MMI and StreamCat data.  The reduced rank method, and covariate transformation and selection procedures will be discussed in subsequent sections. A thorough review of the geostatistical modeling approach discussed in this study is provided in \cite{cressie93spatial} and  \cite{cressie2015spatiotemporal}.

Suppose that $\bm{Y} = (Y(\bm{s}_1), \cdots, Y(\bm{s}_n))'$ is a response vector that is spatially referenced at locations $\bm{s}_i \in D \subset \mathbb{R}^2$. The spatial regression model can be written as
\begin{linenomath}
\begin{align}
\bm{Y} = \bm{X} \bm{\beta} + \bm{z} + \bm{\epsilon}, \label{eq:slm}
\end{align}
\end{linenomath}
where $\bm{X}$ is an $n \times p$ design matrix for the covariates, $\bm{\beta}$ is a $p \times 1$ vector of unknown regression coefficients, $\bm{z}$ is an $n \times 1$ vector of spatially autocorrelated random variables, and $\bm{\epsilon}$ is an $n \times 1$ vector of independent random errors.  The $n \times n$ covariance matrix for the model can be expressed as
\begin{linenomath}
\begin{align}
\bm{\Sigma} = \var(\bm{Y}) = \var{(\bm{z})} + \var{(\bm{\epsilon})} = \bm{R} + \sigma^2_{\epsilon} \bm{I}.  \label{eq:covmtx}
\end{align}
\end{linenomath}
To simplify estimation of (\ref{eq:covmtx}), we assume a stationary covariance function that depends on Euclidean distance and takes an exponential form.  That is, the $(i,j)$ entry of $\bm{R}$ is given by
\begin{linenomath}
\begin{align}
\cov(z(\bm{s}_i), z(\bm{s}_j)) = \sigma^2_z \exp(-\lVert \bm{s}_i - \bm{s}_j \rVert/\alpha), \label{eq:expcov}
\end{align}
\end{linenomath}
where  $\lVert \cdot \rVert$ denotes the Euclidean distance metric, and $\sigma^2_z$ and $\alpha$ are parameters to be estimated. In the geostatistical literature, the parameters $\sigma^2_z$, $\alpha$, and $\sigma^2_\epsilon$ are, respectively, called the partial sill, range, and nugget.  The nugget parameter models residual variation in the response when the separating distance is zero.

Model~(\ref{eq:slm}) is also commonly referred to as the spatial linear model (SLM), and as the universal-kriging model when used for spatial prediction, with the ordinary-kriging model being the special case when $\bm{X}$ is a $n \times 1$ column vector of 1's.  While numerous types of covariance functions have been proposed for the SLM (e.g., \citealp{chiles99geos}, pp.~80--93), we only consider the exponential form in (\ref{eq:expcov}) since estimation of the SLM is computationally demanding with large data sets.  Moreover, for modeling MMI, we focus instead on estimating covariate transformations in $\bm{X}$, which are discussed subsequently in Section~\ref{sec:transformations}.  Regionally varying intercept terms are also included in $\bm{X}$ to account for differences in MMI development in the nine ecoregions.

The parameters of a spatial regression model for a particular data set can be estimated using maximum likelihood (ML; \citealp{cressie93spatial}, p. 92) or restricted maximum likelihood (REML; \citealp{patterson1971reml}; \citealp{harville1974reml}) estimation.  The negative log-likelihood can be expressed as
\begin{linenomath}
\begin{align}
l(\bm{\beta}, \bm{\theta}) =  0.5 \{n \log(2 \pi) + \log(|\bm{\Sigma}(\bm{\theta})|) +
(\bm{Y} - \bm{X}\bm{\beta})' \bm{\Sigma}(\bm{\theta})^{-1} (\bm{Y} - \bm{X}\bm{\beta}) + c\}, \label{eq:ll}
\end{align}
\end{linenomath}
where, for ML, $c=0$, and for REML, $c=-p\log(2\pi) + \log |\bm{X}' \bm{\Sigma}(\bm{\theta})^{-1} \bm{X}|$; the covariance matrix, $\bm{\Sigma}(\bm{\theta})$, is written here to emphasize dependence on the unknown covariance parameters $\bm{\theta}$ (i.e., nugget, partial sill, and range).  Minimizing (\ref{eq:ll}) with respect to $\bm{\beta}$ gives
\begin{linenomath}
\begin{align}
\hat{\bm{\beta}}(\bm{\theta}) = (\bm{X}'\bm{\Sigma}(\bm{\theta})^{-1} \bm{X})^{-1} \bm{X}' \bm{\Sigma}(\bm{\theta})^{-1} \bm{Y}. \label{eq:betahat}
\end{align}
\end{linenomath}
The ML or REML estimators of the covariance parameters, $\hat{\bm{\theta}}$, are obtained by substituting $\hat{\bm{\beta}}(\bm{\theta})$ into $(\ref{eq:ll})$ and minimizing with respect to $\bm{\theta}$; estimators of the regression coefficients are consequently found by substitution of $\hat{\bm{\theta}}$ back into (\ref{eq:betahat}), i.e., $\hat{\bm{\beta}}(\hat{\bm{\theta}})$.   In practice, we obtain ML or REML estimates of $\bm{\theta}$ numerically using the general purpose optimization function \texttt{optim()} provided in the statistical software package R \citep{r}.  Note that REML estimators tend to have less bias and better performance (in terms of mean squared error) than ML estimators, especially when $p$ is large relative to $n$ (\citealp{harville1977ml}; \citealp{cressie93spatial}, p. 93).

Once the parameters are estimated for the spatial regression model, we use universal kriging to make predictions and construct prediction intervals (\citealp{cressie93spatial}, pp. 151--155; \citealp{cressie2015spatiotemporal}, pp. 145--148).  The universal-kriging prediction and variance equations for the response at a new location $\bm{s}_0$ are given by
\begin{linenomath}
\begin{align}
\hat{Y}(\bm{s}_0) &= \bm{x}(\bm{s}_0)' \hat{\bm{\beta}} + \bm{c}(\bm{s}_0)' \bm{\Sigma}^{-1} (\bm{Y} - \bm{X} \hat{\bm{\beta}}) \label{eq:UKpred}\\
\var(\hat{Y}(\bm{s}_0)) &= C(\bm{s}_0,\bm{s}_0) - \bm{c}(\bm{s}_0)' \bm{\Sigma}^{-1} \bm{c}(\bm{s}_0) + \bm{t}(\bm{s}_0)' (\bm{X}' \bm{\Sigma}^{-1} \bm{X})^{-1} \bm{t}(\bm{s}_0), \label{eq:UKvar}
\end{align}
\end{linenomath}
where  $\bm{t}(\bm{s}_0) = \bm{x}(\bm{s}_0) - \bm{X}' \bm{\Sigma}^{-1} \bm{c}(\bm{s}_0)$, $\bm{c}(\bm{s}_0) = \cov(Y(\bm{s}_0), \bm{Y}) = (C(\bm{s}_0, \bm{s}_1), \cdots, C(\bm{s}_0, \bm{s}_n))'$, and $\bm{x}(\bm{s}_0)$ is the covariate vector at $\bm{s}_0$. Note that the covariance function is defined here as $C(\bm{u},\bm{v}) = \sigma_z^2 \exp(||\bm{u}-\bm{v}||/\alpha) + I(\bm{u}=\bm{v})\sigma_{\epsilon}^2$ for all locations $\bm{u},\bm{v} \in D$.  Also, note that (\ref{eq:UKpred}) is derived as the homogeneously linear combination of the data, $\bm{\lambda}'\bm{Y}$ where $\bm{\lambda} \in \mathbb{R}^n$, that minimizes the mean-squared-prediction error, $E(Y(\bm{s}_0) - \bm{\lambda}' \bm{Y})^2$, subject to the unbiasedness constraint $E(\bm{\lambda}' \bm{Y}) = E(Y(\bm{s}_0)) = \bm{x}(\bm{s}_0)' \bm{\beta}$;  and (\ref{eq:UKvar}) is the minimized mean-square-prediction error, often referred to as the kriging variance.

\subsection{Reduced Rank Methods} \label{sec:rrmethod}

For data sets with a large number of records, inverting the covariance matrix when optimizing the log-likelihood function (\ref{eq:ll}) can be computationally burdensome.  For example, the motivating 2008/09 NRSA data set contains nearly 2000 records; thus there is a computational cost when estimating a spatial regression model for this data set.  To accelerate parameter estimation for the SLM we consider reduced rank methods \citep{cressie2008frk, banerjee2008gaussian}.  In this section we specify the reduced rank method used in the study; Section~\ref{sec:selection} will subsequently discuss the application of this method to a computationally efficient covariate selection routine.

Consider a set of $r$ knot locations $\{\bm{k}_i : i = 1, \cdots, r \}$, distributed over the same domain as the observed data, such that $r << n$. Instead of modeling the covariance matrix for $\bm{Y}$ in terms of the Euclidean distances between the observed locations, we can alternatively model the covariance matrix in terms of the knot locations as
\begin{linenomath}
\begin{align}
\bm{\Sigma} = \bm{S}\bm{K}^{-1}\bm{S}' + \sigma^2_{\epsilon} \bm{I}, \label{eq:rrcovmtx}
\end{align}
\end{linenomath}
where $\bm{S}$ is an $n \times r$ matrix with $(i,j)$ element $\sigma^2_z \exp (-\lVert \bm{s}_i - \bm{k}_j \rVert / \alpha )$; $\bm{K}$ is an $r \times r$ matrix with $(i,j)$ element $\sigma^2_z \exp (-\lVert \bm{k}_i - \bm{k}_j \rVert / \alpha )$; and $\sigma^2_z$, $\alpha$, and $\sigma^2_{\epsilon}$ are parameters to be estimated.  An advantage of the specification in (\ref{eq:rrcovmtx}) is that application of the well-known Sherman-Morrison-Woodbury formula (see \cite{henderson1981smw} for a review) yields the following decomposition:
\begin{linenomath}
\begin{align}
\bm{\Sigma}^{-1} = \sigma_{\epsilon}^{-2} \left[ \bm{I} - \bm{S}(\sigma_{\epsilon}^2 \bm{K} + \bm{S}\bm{S}')^{-1} \bm{S}' \right].  \label{eq:smw}
\end{align}
\end{linenomath}
Since (\ref{eq:smw}) only involves inverting an $r \times r$ matrix computation speed is greatly improved.

Note that (\ref{eq:rrcovmtx}) can be viewed as the covariance matrix for $\bm{Y}$ in the spatial mixed effects model $\bm{Y} = \bm{X \beta} + \bm{S \gamma} + \bm{\epsilon}$, where $\bm{\gamma}$ is an $r\times 1$ vector of random effects such that $\var(\bm{\gamma})=\bm{K}^{-1}$.  Also, one property of the covariance matrix specified in (\ref{eq:rrcovmtx}) is that if the knots are the observed data locations $\{\bm{s}_i\}$, then $\bm{S} = \bm{K} = \bm{R}$, and consequently (\ref{eq:rrcovmtx}) is equivalent to the full rank covariance matrix defined in (\ref{eq:covmtx}).

\subsection{Covariate Transformations} \label{sec:transformations}
\begin{figure}[htbp]
\centering
\includegraphics[width=0.99\textwidth]{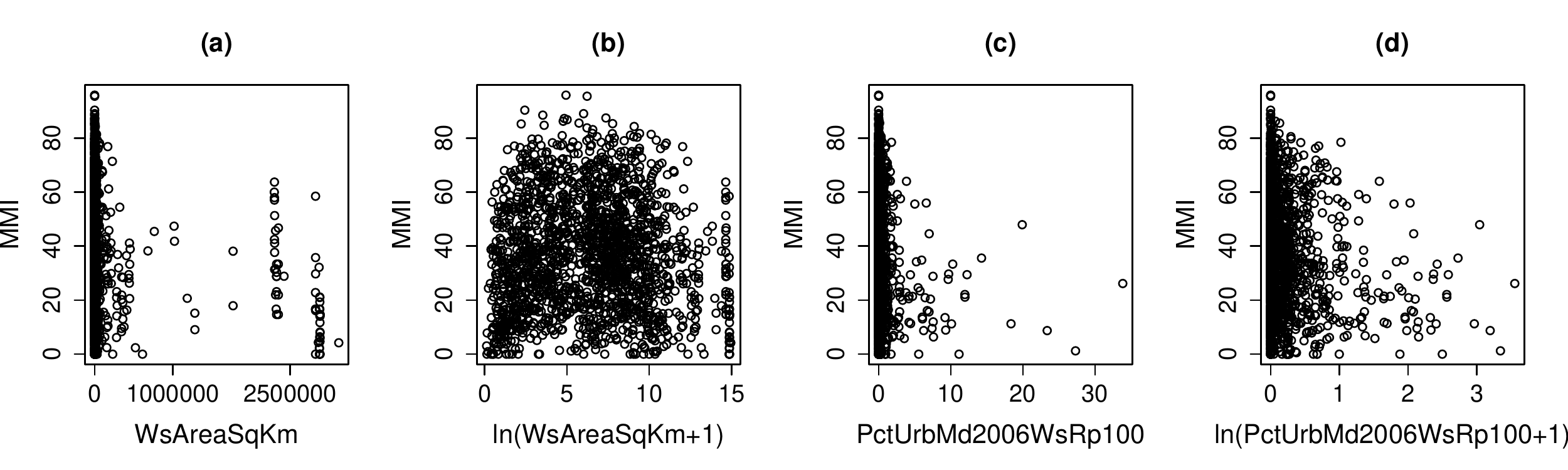}
\caption{Scatter plots of MMI versus the following covariates: (a) watershed area in square kilometers (WsAreaSqKm); (b) natural logarithm  of WsAreaSqKm; (c) percent of watershed area classified as developed, medium intensity land use in 2006 within a 100 meter buffer of a stream reach (PctUrbMd2006WsRp100); (d) natural logarithm of PctUrbMd2006WsRp100.  \label{fig:transf_scatter}}
\end{figure}

Many of the covariates in the StreamCat data set have nonlinear relationships with MMI.  For example, Figure~\ref{fig:transf_scatter}(a,c) shows highly skewed relationships between MMI and two StreamCat covariates: watershed area in square km (WsAreaSqKm); and the percent of watershed area classified as developed, medium intensity land use within a 100-m buffer of a stream reach (PctUrbMd2006WsRp100).  The log-transformation helps linearize the relationship between MMI and PctUrbMd2006WsRp100 (Figure~\ref{fig:transf_scatter}d), and also reveals a quadratic relationship between MMI and WsAreaSqKm (Figure~\ref{fig:transf_scatter}b).  While not shown here, many of the other StreamCat covariates exhibit similar types of nonlinearities, and further motivate considering covariate transformations.

To linearize relationships between the covariates and response variable, we estimate Box-Cox transformations \citep{boxcox1964} for the covariates.  Specifically, we estimate transformations of the form
\begin{linenomath}
\begin{align}
g(x; \lambda_1, \lambda_2) = \label{eq:boxcox}
\begin{cases}
\frac{(x+\lambda_2)^{\lambda_1} - 1}{\lambda_1} & \lambda_1 \neq 0\\
\log(x+\lambda_2) & \lambda_1 = 0,
\end{cases}
\end{align}
\end{linenomath}
where $x > - \lambda_2$.  Note that Box-Cox transformations were first proposed as a way to transform the response variable \citep{boxcox1964}, however, these types of power transformations have also been applied to the independent variables in regression modeling (\citealp{fox2008regression}, pp.~50--63).

Figure~\ref{fig:transf_scatter} suggests that StreamCat covariates can be zero-inflated (e.g., PctUrbMd2006WsRp100), or have quadratic relationships with the response (e.g., log-transformed WsAreaSqKm).  Different types of transformation effects are considered depending on whether or not the StreamCat covariate is zero-inflated.  To estimate transformations for the zero-inflated covariates, we estimate the following linear models for varying values of $\lambda_1$ and $\lambda_2$:
\begin{linenomath}
\begin{align}
y &= \beta_0 + \beta_1 I(x_i \neq 0) + \epsilon, \label{eq:binzero}\\
y &= \beta_0 + \beta_1 g(x_i; \lambda_1,\lambda_2) I(x_i \neq 0) + \epsilon, \label{eq:inter}\\
y &= \beta_0 + \beta_1 I(x_i \neq 0) + \beta_2 g(x_i; \lambda_1,\lambda_2) I(x_i \neq 0) + \epsilon. \label{eq:both}
\end{align}
\end{linenomath}
Here $y$ is MMI, $x_i$ is the $i^{th}$ StreamCat covariate, $\epsilon$ is the error term, and $I(a)$ is the indicator function, equal to one if the argument $a$ is true, and zero otherwise.  For each zero-inflated covariate, we use the Aikaike Information Criterion (AIC; \citealp{akaike1974criterion}) to select optimal $(\lambda_1, \lambda_2)$ and the type of transformation:  zero/nonzero indicator (\ref{eq:binzero}), interaction between the indicator and transformed covariate (\ref{eq:inter}), or both (\ref{eq:both}).  That is, out of several candidate values for $(\lambda_1, \lambda_2)$, we select the Box-Cox parameter values and type of transformation effect that corresponds to the linear model (\ref{eq:binzero}, \ref{eq:inter}, or \ref{eq:both}) with the lowest AIC.  To estimate transformations for the other covariates (not zero-inflated), we estimate the following linear models for varying values of $\lambda_1$ and $\lambda_2$:
\begin{linenomath}
\begin{align}
y &= \beta_0 + \beta_1 g(x_i; \lambda_1,\lambda_2) + \epsilon, \label{eq:linear}\\
y &= \beta_0 + \beta_1 g(x_i; \lambda_1,\lambda_2) + \beta_2 (g(x_i; \lambda_1,\lambda_2))^2 + \epsilon. \label{eq:quadratic}
\end{align}
\end{linenomath}
For each covariate that is not zero-inflated, we again use the AIC to select optimal $(\lambda_1, \lambda_2)$ and the type of transformation: linear (\ref{eq:linear}) or quadratic (\ref{eq:quadratic}) polynomial.  In practice, we vary the values of the exponent parameter $\lambda_1$ between 0 and 3, and try several shifting parameters $\lambda_2$ to ensure that (\ref{eq:boxcox}) is well defined.  We also define a covariate $x_i$ as being zero-inflated if the proportions of zeros is greater than 2\%.  Note that the transformations are estimated separately for each covariate, and that spatial autocorrelation is ignored while estimating transformations because otherwise the procedure would be too slow computationally.

Once transformations have been selected, a new design matrix containing the transformations for each StreamCat covariate can be used to fit either a multiple regression or spatial regression model.  Since we include indicator variables for zero-inflated covariates and quadratic polynomial effects, the transformed design matrix is larger than the design matrix without transformations.  In the next section, we discuss a covariate selection approach for reducing the number of parameters in a spatial regression model.

\subsection{Covariate Selection} \label{sec:selection}

Covariate selection for the spatial regression model is implemented in two phases.  For the first phase, we fit a multiple linear regression model (LM) with the full set of covariates from the StreamCat data set.  Dummy variables for the ecoregions (Figure~\ref{fig:nrsa_map}) are also included as additional covariates in the LM to account for regional variations in MMI development.   Variables are then selected using a backwards stepwise algorithm (i.e., the \texttt{step()} function from \cite{r}).  Note, we start by selecting variables for an LM rather than an SLM since the LM can be rapidly estimated, and software is readily available for variable selection.

For the second phase, we estimate an SLM with the variables selected for the LM.  We then conduct further variable selection for the SLM since some covariates may no longer be significant once spatial autocorrelation is taken into account \citep{verhoef01slm, hoeting2006modelsel}.  Specifically, we repeatedly remove the covariate with the largest absolute t-statistic and re-fit the SLM until the model's AIC score starts to increase.  To speed up ML estimation of the SLM during this procedure we use the reduced rank method (Section~\ref{sec:rrmethod}).  Once all variables are selected for the SLM, REML with the full rank covariance matrix is used to estimate the final model.  A step-by-step description of the covariate selection procedure is provided in the Supplement (Section~S1).

\section{Random Forest Model}

Random forest (RF) modeling has become a popular technique for regression and classification with complex environmental data sets \citep{prasad2006, cutler2007ecology, carlisle2009, evans2011, hill2013, freeman2015tuning}.    In contrast to multiple regression, RF is an entirely nonparametric and algorithmic procedure which makes no \emph{a priori} assumptions about the relationship between the predictor variables and the response.  RF has a reputation for good predictive performance when the data contain a large number predictor variables, and when there are complex nonlinearities and interaction effects in the relationship between the predictors and response variable \citep{cutler2007ecology, strobl2009trees, biau2012}.  In addition, RF provides several measures of variable importance that allow for interpretation of the fitted model (\citealp{hastie09elements}, p.~593).

An RF model can be defined as a collection of regression trees $\{T_b: b=1,\cdots,B\}$ each built from a bootstrap sample of the data set $\{\bm{Y}, \bm{X}\}$.  When growing each tree $T_b$, at each parent node a subset of $m$ of the $p$ predictor variables are randomly selected, and the best split-point is found among those $m$ variables to form two daughter nodes.  The trees in the RF ensemble are grown deep with no pruning. Bagging trees \citep{breiman1996bagging} are the special case when $m=p$ (i.e., all predictor variables are used as candidates for splitting at each node).  An RF prediction at a new site with predictor values $\bm{x} = (x_1, \cdots, x_p)$ is found by averaging the predictions made by each tree in the ensemble:
\begin{linenomath}
\begin{align*}
\hat{f}(\bm{x}) = \frac{1}{B} \sum_{b=1}^B T_b(\bm{x}).
\end{align*}
\end{linenomath}
RF is also commonly used for classification, in which case $T_b(\bm{x})$ takes on discrete values (e.g., 0/1 for binary classification) and the RF prediction can be defined as the majority vote from the collection of class predictions $\{T_b(\bm{x})\}$.  However, in this study we only consider RF for regression.  Note that the RF algorithm produces individual tree estimators with high variance.  That is, a regression tree fit to different portions of the data can yield very different predictions at an unsampled site.  The main idea behind RF is that averaging over many tree models is a way to reduce variance, and thereby improve predictive performance relative to a single tree model \citep{breiman1996bagging, breiman01rf}.  Moreover, only considering a random subset of $m < p$ predictors at each node has the effect of decorrelating the trees in the ensemble, which can further improve the performance of the RF model relative to bagging.  See \cite{hastie09elements} for a comprehensive review of RF and relevant theory.

We implement RF using the R package \texttt{randomForest} \citep{liaw02randomForest}.  There are two tuning parameters when estimating an RF model with this package: the number of trees $B$, and the number of predictor variables $m$ randomly selected at each node.  However, RF is generally insensitive to choice of tuning parameters, and the defaults perform adequately for most data sets \citep{liaw02randomForest, cutler2007ecology, freeman2015tuning}.  In practice, we use $B=1000$ trees and the default $m=p/3$.  We use more trees than the default value (500) since this is recommended for data sets that have a large number of predictors \citep{liaw02randomForest}.  For the RF model of MMI, we also use the full set of StreamCat predictor variables, as well as an additional categorical predictor for the ecoregions.  We do not perform additional subset selection since the RF algorithm is robust to handling large sets of predictor variables \citep{breiman01rf, strobl2009trees,  biau2012}.

\subsection{Quantile Prediction Intervals} \label{sec:qrf}

Prediction intervals for RF can be computed using quantile regression forests (QRF; \citealp{meinshausen2006quantile}).  While RF provides information on the conditional mean of the response,  QRF instead provides information on the conditional distribution function of the response.  Approximation of the conditional distribution function is useful for making quantile predictions and forming associated prediction intervals.  For instance, a 90\% prediction interval for the response at a new site with predictors $\bm{x}$ can be formed as the 0.05 and 0.95 quantile predictions denoted by $[\hat{Q}_{0.05}(\bm{x}), \hat{Q}_{0.95}(\bm{x})]$.  Here $Q_{\alpha}(\bm{x})$ defines the $\alpha$-quantile, that is, the value of the random response variable $Y$ such that the probability $Y$ is less than $Q_{\alpha}(\bm{x})$, for given $\bm{x}$, is exactly equal to $\alpha$; $\hat{Q}_{\alpha}(\bm{x})$ denotes the QRF estimator of this quantity.  A main distinction between the RF and QRF algorithms is that RF only keeps track of the mean value of the response data at each leaf (terminal node) of each tree.  QRF, on the other hand, keeps track of all the response data at each leaf of each tree and approximates the full conditional distribution with this additional information.  In practice, we implement the QRF method using the R package \texttt{quantregForest} \citep{quantregForest}, which builds on the \texttt{randomForest} package also used in this study.

\subsection{Random Forest Regression Kriging}
Random forest regression kriging (RFRK) has been proposed in a number of studies as a way to account for spatial autocorrelation in RF modeling (e.g, \citealp{li2011can, hengl2015mapping, fayad2016regional}).  For RFRK, a prediction at a new site is given by summing the RF prediction and the kriging prediction of the RF residual.  Formally, an RFRK prediction of $Y(\bm{s}_0)$, at a new site $\bm{s}_0$, is given by
\begin{linenomath}
\begin{align*}
\hat{Y}(\bm{s}_0) = \hat{f}_{RF}(\bm{x}(\bm{s}_0)) + \hat{e}(\bm{s}_0)
\end{align*}
\end{linenomath}
where $\hat{f}_{RF}$ is the RF prediction with covariates $\bm{x}(\bm{s}_0)$, and $\hat{e}(\bm{s}_0)$ is the kriging prediction of the residual.  In practice, we use ML to estimate a simple-kriging model for the RF residuals that assumes a zero mean (i.e, $E(e(\bm{s}))=0$) and exponential model for the covariance matrix.  Additionally, we use the simple-kriging variances for the residuals to construct prediction intervals.  Computational details are provided in  the Supplement (Section~S2).

\section{Performance Measures} \label{sec:measures}
We assess the performance of different models for MMI using 10-fold cross validation.  Seven models of MMI are considered for the comparison: (1) an ordinary-kriging model, i.e., an SLM with no covariates and a single intercept term ($\bm{X}$ is a $n\times1$ column vector of 1's); (2) an LM with no transformations; (3) an SLM with no transformations; (4) an LM with transformations; (5) an SLM with transformations; (6) an RF model; and (7) an RFRK model.

The root-mean-square prediction error (RMSPE) and coverage of the prediction intervals are used to evaluate the cross-validation performance of the different models.  Let $Y_i$ denote the $i^{th}$ observed value and $\hat{Y}_i$ the 10-fold cross-validation prediction.  The RMSPE is computed as the square root of the mean of $(Y_i-\hat{Y}_i)^2$ and coverage of the 90\% prediction intervals is computed as the mean of $I(\hat{Y}_i - 1.645 \se(\hat{Y}_i) < Y_i < \hat{Y}_i + 1.645 \se(\hat{Y}_i))$ for all $i$ in $1,\cdots,n$, where $I$ is the indicator function and $\se(\hat{Y}_i)$ is the prediction standard error.  Since we estimate quantile prediction intervals for RF, the coverage of the 90\% prediction intervals is computed as the mean of $I(\hat{Q}_{0.05}(\bm{x}_i) < Y_i < \hat{Q}_{0.95}(\bm{x}_i))$ for all $i$ in $1,\cdots,n$, where $\bm{x}_i$ are the predictor variables at the $i^{th}$ site and $\hat{Q}_{\alpha}(\bm{x}_i)$ is the 10-fold cross-validation prediction of the $\alpha$-quantile defined in Section~\ref{sec:qrf}.

\section{Results} \label{sec:results}
The 10-fold cross-validation performance measures for modeling MMI are presented in Table~\ref{tab:cv_metrics}.  In terms of RMSPE, the SLM with transformations, respectively followed by RFRK, RF, and the LM with transformations resulted in the best performance.  The SLM and LM without transformations did not perform nearly as well; and not surprisingly, the ordinary-kriging model had the lowest RMSPE.  The results show that covariate transformations for the LM and SLM were necessary to obtain performance comparable with RF.  The covariate transformations also resulted in an LM which had approximately the same performance as RF.  Accounting for spatial autocorrelation improved the performance of the SLM relative to the LM for both the transformed and untransformed cases.  Moreover, additional covariate selection for the SLM (Section~\ref{sec:transformations}) resulted in a more parsimonious model than the LM.  The RFRK model also performed better than the RF model, although the difference was not substantial (Pearson correlation between RF and RFRK cross-validation predictions was greater than 0.98).

\begin{figure}[htbp]
\centering
\includegraphics[width=0.6\textwidth]{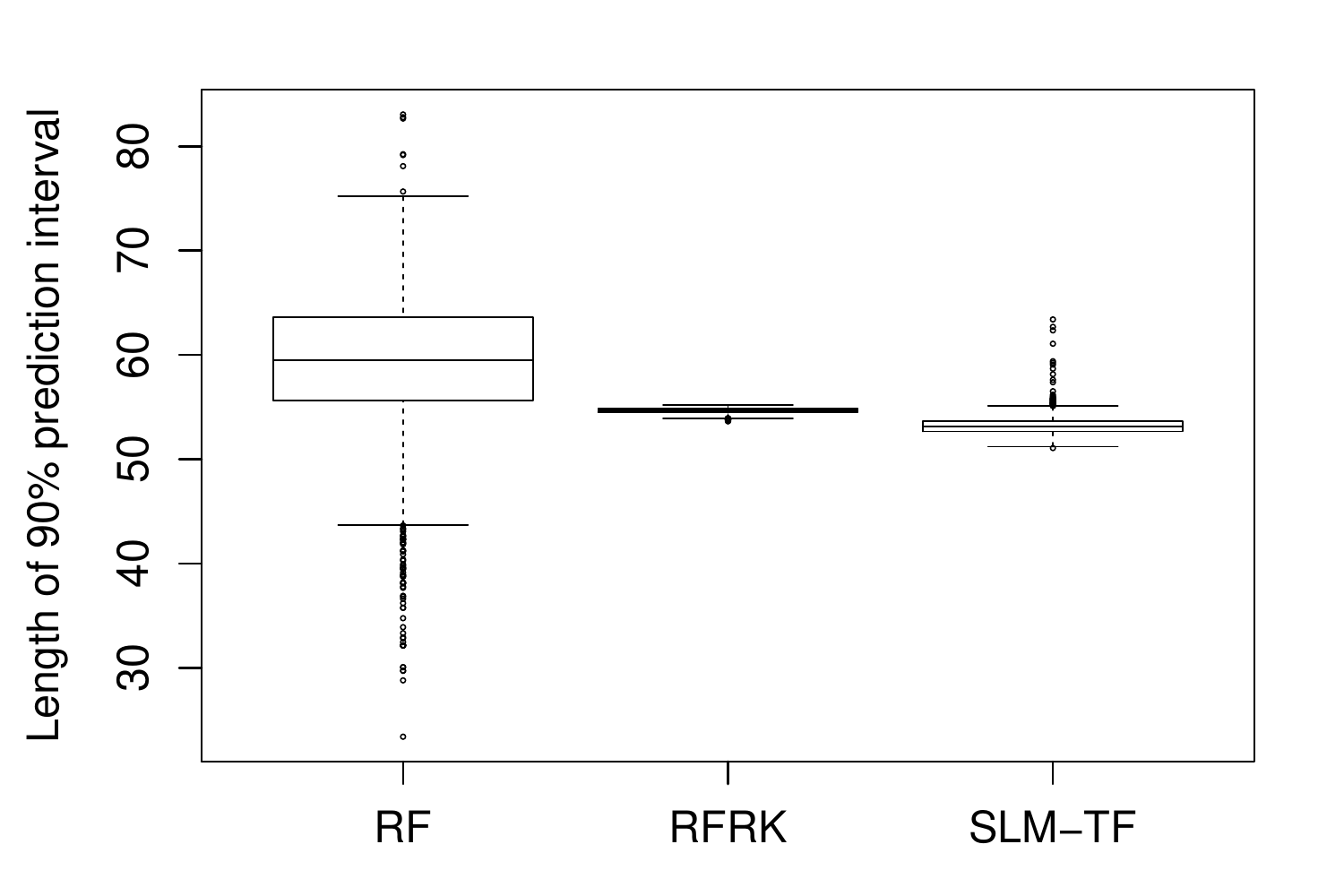}
\caption{Boxplots of the lengths of 90\% predictions intervals (from cross validation) for the RF model, RFRK model, and SLM with transformations (SLM-TF).  Prediction intervals for RF were computed using the quantile regression forest method, while prediction intervals for RFRK and SLM were computed using the kriging variances. \label{fig:boxplot90PI}}
\end{figure}

The coverages of the 90\% and 95\% prediction intervals were close to nominal for all models in Table~\ref{tab:cv_metrics}.  That is, for each method, the prediction intervals computed during cross-validation contained the true observed MMI values with approximately the correct proportion.  However, even though the coverages were similar, there were considerable differences between the lengths of the prediction intervals from the different methods (Figure~\ref{fig:boxplot90PI}).  For instance, there was much greater variability in the lengths of the RF quantile prediction intervals than the SLM and RFRK prediction intervals.  The lengths of the prediction intervals for the SLM also showed a positive skew (Figure~\ref{fig:boxplot90PI}); this can be explained by the universal-kriging variances getting larger for sites which fall away from than bulk of the data in the covariate space \citep{hengl2004rk}.  The distribution of the prediction interval lengths for RFRK was more symmetric, in comparison, since the simple-kriging variances only account for the uncertainty due to relative geographic distances between points, and not possible quantitative extrapolation in the covariates.

\begin{figure}[htbp]
\centering
\includegraphics[width=0.9\textwidth]{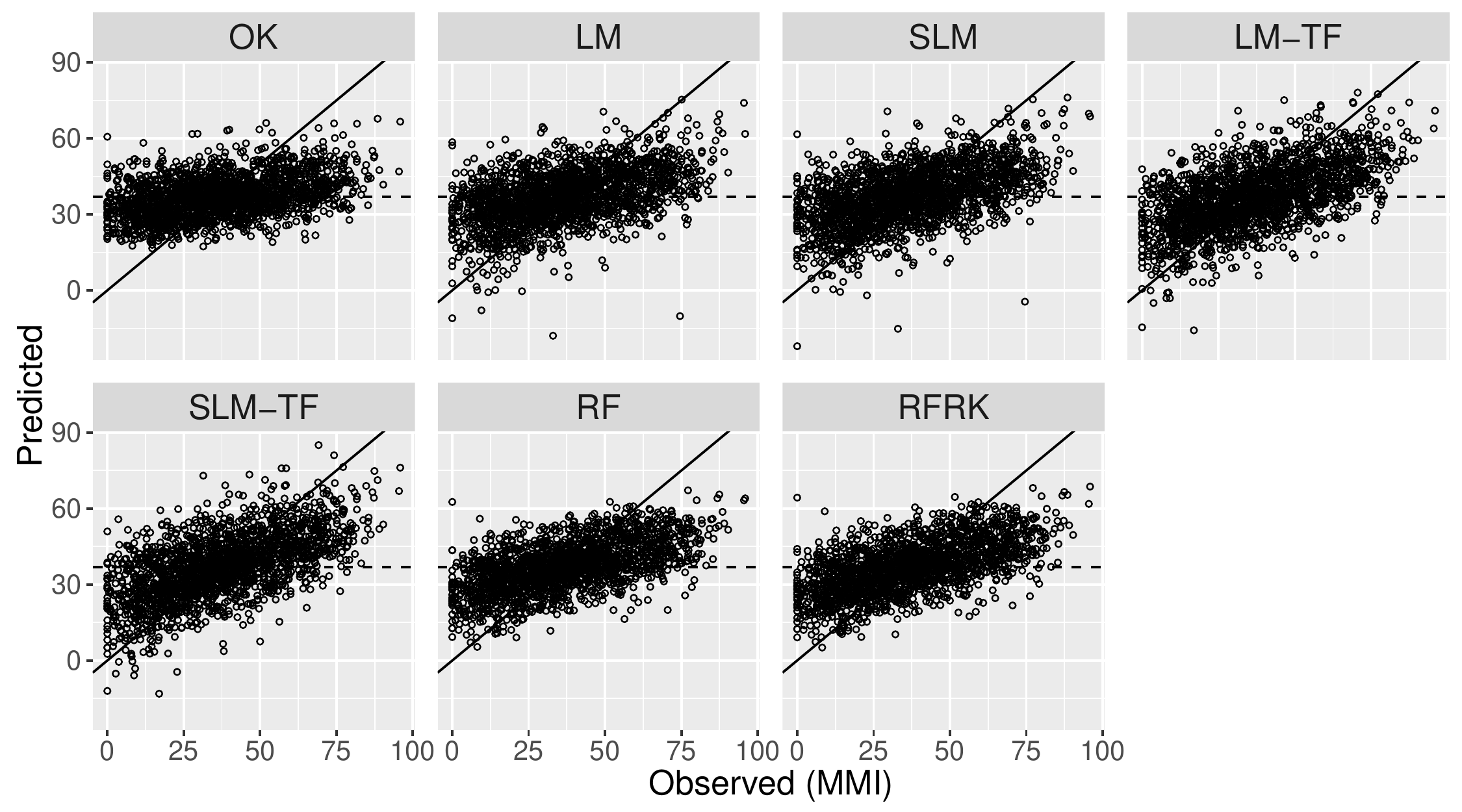}
\caption{Scatter plots of predicted versus observed MMI values; predictions are from 10-fold cross-validation.  The 1-1 line (solid) and mean observed MMI value (36.9; dashed horizontal line) are also shown in each panel. Labels for the seven models are defined in Table~\ref{tab:cv_metrics} and the text. \label{fig:pred_obsv}}
\end{figure}

Scatter plots of predicted versus observed values are presented in Figure~\ref{fig:pred_obsv}.  The scatter plots reveal that the predictions from the different models tend towards the mean of the observed MMIs (36.9).  This effect was most pronounced for the ordinary-kriging, RF, and RFRK models; in comparison, the LM and SLM had wider distributions of predicted values.   While most of the predictions from the LM and SLM (with and without transformations) were within the defined range of the MMI (0--100), Figure~\ref{fig:pred_obsv} shows that a small percentage ($< 0.5\%$) of predictions were negative.  The predictions from the RF and RFRK models, on the other hand, were contained within the observed MMI range.  Note that, by definition, RF models cannot predict outside the range of the observed data since each tree model within the forest makes predictions by taking the mean of the response data falling within a given leaf (terminal) node.

Covariance parameter estimates (nugget, partial sill, and range) for the spatial regression models are presented in Table~\ref{tab:covest}.  The ordinary kriging model has a larger estimated nugget parameter, $\hat{\sigma}^2_{\epsilon}$, and smaller nugget-to-sill ratio, $\hat{\sigma}^2_{\epsilon} / (\hat{\sigma}^2_{z} + \hat{\sigma}^2_{\epsilon})$, than the SLM.  This is expected since the covariates in the SLM explain additional variation in MMI not accounted for by ordinary kriging.  Moreover, the spatially-referenced StreamCat covariates in the SLM account for some spatial autocorrelation in MMI.  To further assist interpretation, Table~\ref{tab:covest} also presents the effective range, $-\hat{\alpha} \log[0.01 * (\hat{\sigma}_z^2 + \hat{\sigma}_{\epsilon}^2) / \hat{\sigma}^2_z]$, which is defined here as the distance beyond which spatial autocorrelation is less than 0.01 (i.e., the distance $h$ found by solving $\rho(h) = C(h) / C(0) = 0.01$; \citealp{irvine2007spatial}).  The effective ranges for ordinary-kriging and the SLM reveal that spatial autocorrelation in the data is close to zero for distances beyond 480--580km.  Additionally, for the RFRK model, both the effective range (160km) and nugget-to-sill ratio (0.951) indicate little, perhaps negligible, spatial autocorrelation in the RF residuals.  Note that, for the effective range calculations we use 0.01 instead of 0.05, which is more common, since the nugget-to-sill ratio for RFRK is greater that 0.95.

\begin{figure}[htbp]
\centering
\includegraphics[width=0.95\textwidth]{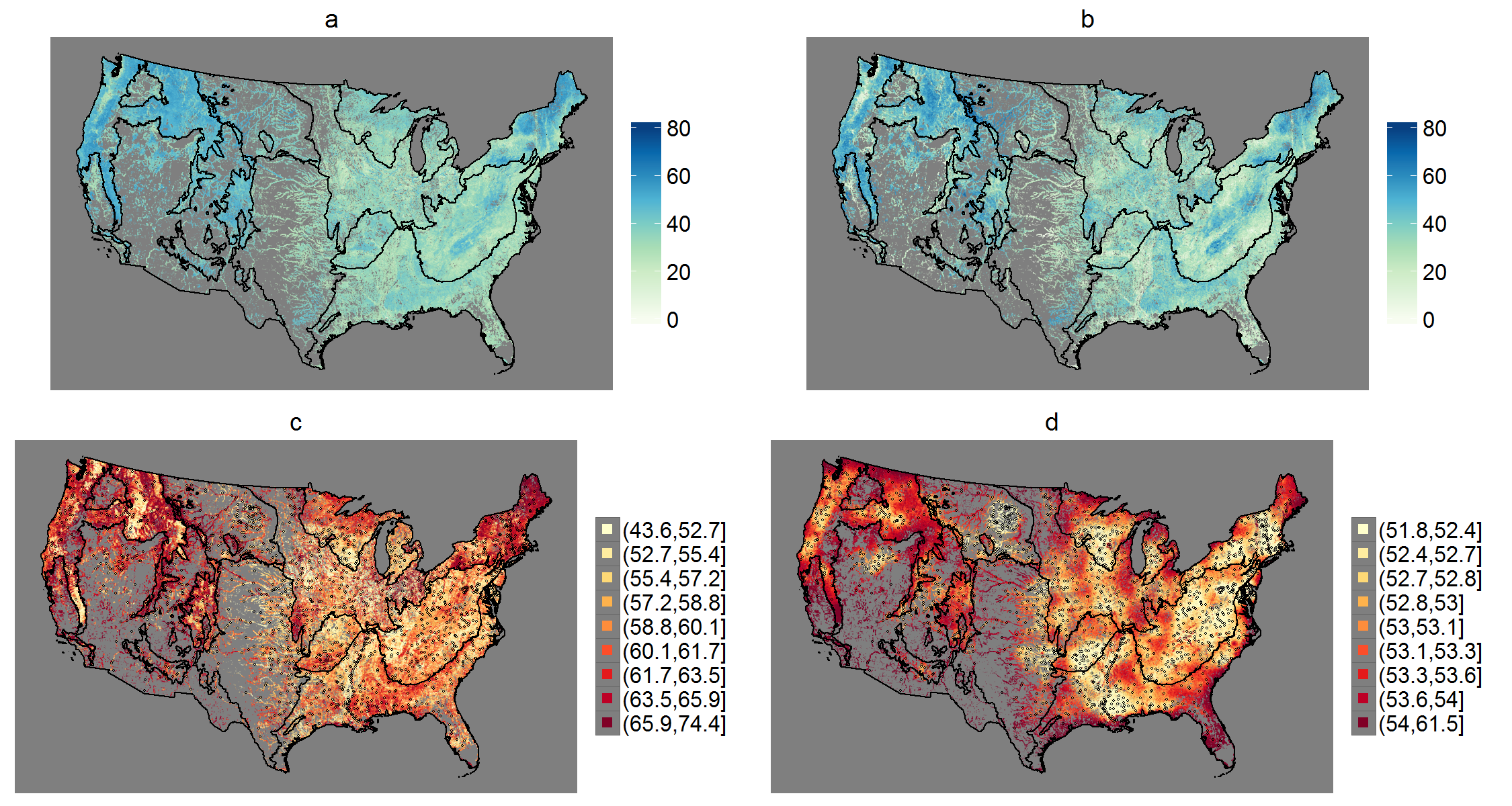}
\caption{Maps of MMI predictions and lengths of 90\% prediction intervals for RF (a,c) and SLM with transformations (b,d). The prediction sites are the 1.1 million perennial stream reaches (catchments) in the NRSA sampling frame.  Note the different scales in the maps of prediction interval lengths (c,d).  Also note that SLM predictions (b) were truncated at the 0.005 and 0.995 quantiles, and negative predictions ($<1.3$\% of sites) were set to zero since MMI is defined between 0--100. \label{fig:predmaps}}
\end{figure}

Maps of the RF and SLM predictions of MMI are presented in Figure~\ref{fig:predmaps} (a,b).  The maps show MMI predictions for 1.1 million perennial stream reaches within the CONUS.  Again, predictions were made only for perennial stream reaches since the sampling frame for the 2008/09 NRSA is limited to these types of streams.  Overall, the maps show similar spatial patterns in the MMI predictions from the two models.  As expected, regions dominated by urban or agricultural land tend to have lower MMI predictions than more remote regions.  The most noticeable difference between the prediction maps is that the SLM shows a wider distribution of predicted values than RF, with sharper differences between regions with low and high MMI predictions (e.g., Willamette Valley versus Cascades in Oregon; Piedmont versus Blue Ridge Mountains in Georgia, S. and N. Carolina, and Virginia).  Also note that the RF predictions never reach zero, while about 1.2\% of the SLM predictions are negative and set to zero in Figure~\ref{fig:predmaps} since MMI is defined between 0--100.

In contrast to the prediction maps, the maps of the RF and SLM prediction errors (lengths of 90\% prediction intervals) are strikingly different (Figure~\ref{fig:predmaps}c,d).  There is much greater variability in the lengths of prediction intervals in the map for the RF model than the SLM (note, this is consistent with the cross-validation results in Figure~\ref{fig:boxplot90PI}).  Moreover, the SLM prediction intervals show greater precision in regions that are more densely sampled (e.g., Mississippi Basin).  Contrastingly, the precision of the RF quantile prediction intervals do not apparently scale with the density of sampling locations.

Lastly, both the SLM and RF models provide measures of variable importance.  For the SLM, covariates can be ranked in terms of the absolute t-statistics for the coefficients.  For RF, covariates can be ranked in terms of a permutation-based measure (i.e., the average increase in mean-square error when each covariate is permuted in the out-of-bag data; \citealp{hastie09elements}, p. 593).  The highest ranked variables for modeling MMI were similar for the RF model and SLM with transformations.  Specifically, the covariates for watershed area, average topographic wetness index, and ecoregions were ranked in the top five for both models.  An SLM regression coefficient summary and RF variable importance plot are provided in the Supplement (Section~S3).

\section{Simulation} \label{sec:simulation}
Modeling the MMI data indicated that the SLM, with appropriate transformations of the covariates to obtain near linear relationships between response and covariates, performed slightly better than RF. However, this is but a single data set with apparently little autocorrelation in the residuals. We explore the affect of nonlinear relationships between response and covariates, $R^2$, and varying amounts of autocorrelation, through simulations.

Data for this simulation study are generated from the following model:
\begin{linenomath}
\begin{align}
y(\bm{s}) &= f(x_1, \cdots, x_4) + \delta(\bm{s}) \nonumber = c[a g(x_1,x_2) +  h(x_3, x_4)] + \delta(\bm{s}) \nonumber\\
&= c[a \sin(5\pi x_1 x_2) + 2 x_3 - x_4] + \delta(\bm{s}). \label{eq:sim}
\end{align}
\end{linenomath}
Here $\delta(\bm{s}) = z(\bm{s}) + \epsilon$ is a spatially autocorrelated error term such that $\cov(z(\bm{s}), z(\bm{s}+\bm{h})) = \sigma^2_z \exp(-\lVert \bm{h} \rVert/\alpha)$ and $\var(\epsilon) = \sigma^2_{\epsilon}$ is the nugget effect.  The parameter $c$ governs the proportion of variance in $y$ explained by the covariates in the systematic component of the model $f$.  The parameter $a$ governs amount of nonlinearity in $f$, which is decomposed into a nonlinear term $g$ and linear term $h$.  Note that the sine function, with multiplicative interaction between $x_1$ and $x_2$, in (\ref{eq:sim}) was chosen since it is difficult to recover with a linear model, and so RF is expected to have advantages if the data are generated from this type of nonlinear function.

The following characteristics of the simulated data are varied by adjusting the values of the parameters in (\ref{eq:sim}):
\begin{itemize}
\item The amount of spatial autocorrelation in the error term $\delta(\bm{s})$.  We set $\sigma^2_z = 1$ and $\sigma^2_{\epsilon} = 9$ for a low amount of autocorrelation, and $\sigma^2_z = 9$ and $\sigma^2_{\epsilon} = 1$ for a high amount of autocorrelation.  The range parameter is always $\alpha = 0.5$.
\item The empirical $R^2$, i.e., the proportion of variation in $y$ explained by $f$.  The value of parameter $c$ is adjusted in each simulation run to give an empirical $R^2$ which is either high (0.9) or low (0.1).
\item Whether the linear or nonlinear term dominates.  The value of parameter $a$ is adjusted in each simulation run so that the proportion of variance in $f$ explained by the nonlinear term $g$ is either high (0.9) or low (0.1).
\end{itemize}
This gives a total of $2^3=8$ cases since there are 2 levels (high/low) for each characteristic (spatial autocorrelation, empirical $R^2$, and amount of nonlinearity).  The 8 cases are summarized in Table~\ref{tab:sim}.

For each simulation case, we generated 20 data sets $[\bm{y}; \bm{x}_1, \cdots, \bm{x}_4]$ of 1500 points with locations randomly generated over the unit square.  For each data set, 500 points were used for training, and the other 1000 as a test set.  Values for the covariates $\bm{x}_i$ were drawn from Unif[0,1].  Data from the spatially autocorrelated error term $\delta(\bm{s})$ in (\ref{eq:sim}) were generated using the Cholesky decomposition method (\citealp{cressie93spatial}, pp.~201--202). Values of parameters $c$ and $a$ were selected to fix the empirical $R^2$ and amount of nonlinearity in the simulated data sets generated for each case.  Since values of $c$ and $a$ varied, Table~\ref{tab:sim} presents the averaged values.

The LM, SLM, RF, and RFRK models are compared in the simulations.  The SLM is fit using REML with the full rank covariance matrix, and no covariate transformations are considered. Model performance measures (RMSPE and prediction interval coverage) are averaged over the 20 simulated data sets generated for each case.  The simulation code is provided in an R package available at \url{https://github.com/ericwfox/slmrf}.

\subsection{Simulation Results}

Simulation results are presented in Table~\ref{tab:sim}. When there was a high amount of autocorrelation in the error term and $R^2=0.1$ (case 2,6), SLM and RFRK performed substantially better than LM and RF.  When $R^2=0.9$ and the nonlinear component dominated (case 3,4), RF and RFRK performed substantially better than LM and SLM; RFRK also performed better than RF when there was a high amount of autocorrelation in the error term (case 4).  When $R^2=0.9$ and the linear component dominated (case 7,8), the SLM had the best performance among all methods; RFRK also performed better than LM when there was a high amount of autocorrelation in the error term (case 8). When the nugget effect dominated (case 1,5), all models performed similarly in terms of RMSPE, and the SLM had slightly better performance than other methods.  For all cases, SLM performed better than LM, and RFRK performed at least as well as RF. However, this is reasonable since there was some amount of autocorrelation in the data generated for each case.  Moreover, when there was only a small amount of autocorrelation in the error term and $R^2=0.9$ (case 3,7), the spatial models performed approximately as well as the non-spatial models (i.e., RF performed as well as RFRK in case 3, and LM performed approximately as well as SLM in case 7).

The coverages of the 90\% prediction intervals for LM, SLM, and RFRK were close to nominal for all simulation cases.  The quantile prediction intervals for RF showed over-coverage when the linear term dominated and $R^2=0.9$ (case 7,8), but were otherwise reasonable, although not as precise as the other methods.

\section{Discussion and Conclusions}
In this article we compared spatial regression and RF methods for modeling stream condition (MMI) with over 200 potential covariates.  We used the models for prediction and uncertainty quantification of MMI at 1.1 million perennial stream reaches across the CONUS.  Initial exploratory analysis revealed highly nonlinear relationships between StreamCat covariates and MMI scores, which motived the application of Box-Cox transformations for the covariates.  We found that the SLM with transformations and RF model performed comparably well in terms of cross-validated RMSPE.  However, the SLM without transformations did not perform nearly as well as RF.  Thus, the estimation of transformations to account for nonlinearities in the data was an important step for constructing a suitable spatial regression model.  The maps of the SLM and RF predictions showed similar spatial trends in stream condition, although the RF predictions were smoother and had greater tendency to concentrate around the mean.  Many of the top predictors identified by each method were also similar.

A novel contribution of this study was the assessment and comparison of prediction intervals for the spatial regression and RF methods. The construction of prediction intervals is not yet common practice in RF modeling.  In contrast to geostatistics, there is no consensus in the RF literature on best practices for uncertainty quantification.  We investigated two ways to construct prediction intervals for RF models: first, by using the quantile regression forest method of \cite{meinshausen2006quantile}; and second, by fitting the RFRK model and using the simple-kriging variances for the RF residuals to form intervals.  We found that coverages of the prediction intervals for the SLM, RF, and RFRK models of stream condition were close to nominal (Table~\ref{tab:cv_metrics}).  However, the lengths of the RF prediction intervals, computed using quantile regression forests, had much greater variability than the SLM and RFRK prediction intervals.  One explanation for these differences is that the kriging variances are optimized by minimizing the mean-square-prediction error, whereas the RF quantile prediction intervals are not found by directly minimizing a loss function.  The large amount of variability in the prediction interval lengths for quantile regression forests was also acknowledged in \cite{meinshausen2006quantile} in applications to a variety of data sets.

The results of this study indicate that there are several trade-offs when deciding between an RF or spatial regression approach to modeling a large environmental data set.  We summarize these below:
\begin{itemize}
\item  Predictions from RF will always be within the range of the observed data, whereas spatial regression can extrapolate outside this range.  In the context of modeling MMI, this was an advantage of the RF approach since the MMI is bounded between 0-100; the SLM also generated some negative MMI predictions that needed to be set to zero in the prediction map (Figure~\ref{fig:predmaps}).  However, in applications to other data sets, predicting within the range of sampled values is not, in general, an advantage.  For instance,  for an unbounded normally distributed response variable, if only 1\% of the data were sampled, then we would expect that, in the other 99\% unsampled sites, there will be values both greater and less than the values in the sample.
\item The SLM stands out as a better descriptive model for ecological processes than RF.  The initial exploratory analysis and transformation procedure provided insights into the functional relationships between the covariates and MMI response variable.  In contrast, the ways in which RF deals with nonlinearities and interactions are hidden in the ensemble of trees and difficult to interpret.
\item RF has a slight computational edge over the SLM.  The RF algorithm is easy to implement using the \texttt{randomForest} package, and RF models are generally insensitive to values of the tuning parameters.  However, computational considerations for fitting an SLM are also not overly-demanding with modern approaches such as REML and reduced rank methods.
\item The results from the data analysis (Section~\ref{sec:results}) suggest advantages to the spatial regression approach for uncertainty quantification.  The SLM prediction intervals were narrower, on average, than the RF quantile prediction intervals.  Moreover, the prediction intervals for the SLM are more suitable for spatial data since they scale with sampling density.
\end{itemize}
Generally, in applications to similar types of large environmental data sets, RF has considerable advantages over the SLM only when treating each approach as a black box method, as we discussed in Section~\ref{sec:lit}.  We recommend the SLM if the practitioner takes a more careful approach to modeling by exploring the data, estimating covariate transformations to account for nonlinearities, and implementing model diagnostics.  While RF is easier to implement, the SLM has advantages for inferential tasks such as constructing prediction intervals and assessing covariate significance.

The simulations demonstrated that no single type of model performs best under all conditions, and that each method is designed for specific purposes.  For data sets with a high amount of nonlinearity in the covariates, and transformations to linearity are difficult or impossible, RF or RFRK are superior methods that can uncover patterns and interactions that are difficult to recover with an LM or SLM.  However, for data sets with a high amount of spatial autocorrelation and linear structure in the covariates, the SLM is the superior method.  Also, if the data have a small amount spatial autocorrelation, then it may not be worth the computational effort to fit a spatial model, and either RF or LM are sufficient.  The simulation results also indicate that the SLM prediction intervals generally have better coverage than the RF quantile prediction intervals.

Going back to our introductory paragraph, environmental data may be large in $n$ (rows) or $p$ (columns). For spatial models, research for large $n$ is very active, including the reduced rank approaches, among others. Less attention is given to large $p$, although many proven techniques can be combined into an overall strategy. We suggest that modelers of spatial data carefully consider transformations to linearity and subsequent removal of covariates to obtain a parsimonious set of (transformed) covariates, including possible interactions. We provided one such example, creating indicator variables for covariates with excessive zeros, using Box-Cox transformations on nonzero covariate values, and creating their interaction. Model selection was possible for large $p$ in the presence of large $n$ using reduced rank methods. We stress that this is not the only strategy, but rather an example of how to proceed for both large $n$ and $p$, which has received little attention. Given such a strategy, we created an SLM model that slightly outperformed RF, which requires less interaction with the data.  Conclusively, there is no correct way to statistically analyze and model large environmental data sets.  The results of this study suggest that a variety of modeling approaches can be considered, and that each approach can lead to different insights into the data set and applied problem.  By comparing spatial regression to RF we ultimately found ways to improve both techniques.

\clearpage

\section*{Acknowledgments}
\noindent We thank Dave Holland (US EPA, National Exposure Research Laboratory, Exposure Methods and Measurements Division) for providing valuable comments that improved this paper.  This manuscript has been subjected to review by the Western Ecology Division of ORD's National Health and Environmental Effects Research Laboratory and approved for publication.  Approval does not signify that the contents reflect the views of the Agency, nor does mention of trade names or commercial products constitute endorsement or recommendation for use.  The data from the 2008/09 NRSA used in this paper resulted from the collective efforts of dedicated field crews, laboratory staff, data management and quality control staff, analysts and many others from EPA, states, tribes, federal agencies, universities and other organizations. For questions about these data, please contact \url{nars-hq@epa.gov}.

\section*{Supplementary Material}
Supplementary material related to this manuscript can be accessed at the following GitHub repository: \url{https://github.com/ericwfox/slmrf}.  The repository contains an R package with the code for the simulation study in Section~\ref{sec:simulation}.  The repository also contains a supplementary PDF with (1) a step-by-step description of the covariate selection procedure; (2) computation details of the random forest regression kriging computations; and (3) additional tables and figures.

\bibliographystyle{apalike}
\bibliography{fox2018_SLM_RF}
\clearpage

%
%
\begin{table}[ht]
\centering
\caption{Cross-validation performance results for models of MMI. \label{tab:cv_metrics}}
\begin{tabular}{lrrrr}
\hline
\hline
Model & k & RMSPE & PIC90 & PIC95 \\
\hline
OK & 4 & 18.55 & 0.900 & 0.962 \\
LM & 49 & 17.70 & 0.898 & 0.960 \\
SLM & 37 & 17.17 & 0.895 & 0.953 \\
LM-TF & 67 & 16.55 & 0.901 & 0.957 \\
SLM-TF & 51 & 16.13 & 0.893 & 0.955 \\
RF &  & 16.52 & 0.914 & 0.955 \\
RFRK &  & 16.37 & 0.907 & 0.959\\
\hline
\end{tabular}
\vspace{1ex}\\
\raggedright
NOTE: $k$ is the number of parameters, PIC90 and PIC95 are the coverages of the 90\% and 95\% prediction intervals, OK is the ordinary-kriging model, LM-TF is the LM with transformations, SLM-TF is the SLM with transformations, and other abbreviations are defined in the text.
\end{table}

\begin{table}[ht]
\centering
\caption{Estimated covariance parameters.\label{tab:covest}}
\begin{tabular}{rrrrr}
\hline
& OK & SLM & SLM-TF & RFRK \\
\hline
Nugget & 278.08 & 257.17 & 226.78 & 261.08 \\
Partial Sill & 135.05 & 68.59 & 53.03 & 13.52 \\
Range & 139.09 & 189.31 & 167.98 & 100.66 \\
Effective Range & 485.03 & 576.87 & 494.19 & 160.44 \\
Nugget-to-Sill Ratio & 0.67 & 0.79 & 0.81 & 0.95 \\
\hline
\end{tabular}
\vspace{1ex}\\
\raggedright
NOTE: The effective range is the distance (km) beyond which spatial autocorrelation is less than 0.01, and the nugget-to-sill ratio is given by  $\hat{\sigma}^2_{\epsilon} / (\hat{\sigma}^2_{z} + \hat{\sigma}^2_{\epsilon})$.
\end{table}

\begin{table}[ht]
\centering
\caption{Simulation results.\label{tab:sim}}
\begin{tabular}{rrrrlrrrrrrrrrr}
\hline
\hline
\multicolumn{7}{c}{} &
\multicolumn{2}{c}{LM} &
\multicolumn{2}{c}{SLM} &
\multicolumn{2}{c}{RF} &
\multicolumn{2}{c}{RFRK}\\
\cline{8-15}
 & & $R^2$ & $\sigma^2_{\epsilon}$ & $\sigma^2_{z}$ & a & c & \scriptsize{RMSPE} & \scriptsize{PIC90} & \scriptsize{RMSPE} & \scriptsize{PIC90} & \scriptsize{RMSPE} & \scriptsize{PIC90} & \footnotesize{RMSPE} & \scriptsize{PIC90}\\
\hline
1 & NL & 0.1 & 9 & 1 & 2.94 & 0.51 & 3.28 & 0.897 & \textbf{3.22} & 0.893 & 3.32 & 0.869 & 3.26 & 0.892 \\
2 & NL & 0.1 & 1 & 9 & 2.94 & 0.43 & 2.77 & 0.901 & \textbf{1.57} & 0.894 & 2.80 & 0.874 & 1.64 & 0.897 \\
3 & NL & 0.9 & 9 & 1 & 2.94 & 4.56 & 9.03 & 0.916 & 9.02 & 0.914 & \textbf{7.45} & 0.914 & \textbf{7.45} & 0.897 \\
4 & NL & 0.9 & 1 & 9 & 2.94 & 3.87 & 7.65 & 0.914 & 7.40 & 0.917 & 6.29 & 0.916 & \textbf{5.96} & 0.894 \\
5 & L & 0.1 & 9 & 1 & 0.33 & 1.52 & 3.16 & 0.897 & \textbf{3.09} & 0.892 & 3.24 & 0.866 & 3.18 & 0.892 \\
6 & L & 0.1 & 1 & 9 & 0.33 & 1.29 & 2.66 & 0.903 & \textbf{1.34} & 0.894 & 2.74 & 0.869 & 1.53 & 0.894 \\
7 & L & 0.9 & 9 & 1 & 0.33 & 13.69 & 4.23 & 0.900 & \textbf{4.19} & 0.899 & 4.54 & 0.959 & 4.51 & 0.899 \\
8 & L & 0.9 & 1 & 9 & 0.33 & 11.61 & 3.58 & 0.899 & \textbf{2.82} & 0.907 & 3.85 & 0.955 & 3.16 & 0.905 \\
\hline
\end{tabular}
\vspace{1ex}\\
\raggedright
NOTE:  Details on labels used in table are provided in Section~\ref{sec:simulation}. The first column indicates the case number and the second column indicates whether the linear (L) or nonlinear (NL) structural component of the model (Equation~\ref{eq:sim}) dominates.  Values for parameters $a$ and $c$, RMSPE, and coverage of 90\% prediction intervals were averaged over 20 independent simulation runs.
\end{table}

\clearpage

%
%

\clearpage

\end{document}